\documentclass[conference,letterpaper,final,10pt]{IEEEtran}
\usepackage{setspace,amssymb,graphicx,enumerate,amsfonts,amsmath,color,cite,algorithm,algpseudocode,lipsum}
\usepackage[english]{babel}

\newcommand{\beq}{\begin{equation*}}
\newcommand{\eeq}{\end{equation*}}
\newcommand{\bea}{\begin{eqnarray}}
\newcommand{\eea}{\end{eqnarray}}
\newcommand{\beal}{\begin{align*}}
\newcommand{\eeal}{\end{align*}}
\newcommand{\bei}{\begin{itemize}}
\newcommand{\eei}{\end{itemize}}
\newcommand{\denk}{\begin{equation*} \begin{aligned}}
\newcommand{\denke}{\end{aligned}\end{equation*}}
\def\bary{\begin{array}}
\def\eary{\end{array}}

\begin{document}
\title{Common Rate Maximization in Two-Layer\\Cellular Radio Systems}
\author{\IEEEauthorblockN{Kemal Davaslioglu, Ender Ayanoglu}
\IEEEauthorblockA{Center for Pervasive Communications and Computing\\
Department of Electrical Engineering and Computer Science,
University of California, Irvine}}\maketitle
\author{Kemal Davaslioglu, Ender Ayanoglu}\maketitle
\begin{abstract}
We address the common rate maximization problem in two-layer cellular networks where high-power and low-power base stations are colocated in the same geographical area. Interference becomes a serious problem when two or more layers are considered in the same network. For this purpose, power control in the downlink needs to be used to limit the interference and to fully exploit the benefits of additional layer deployments. We present an analytical framework to the common rate maximization problem both with and without maximum power constraints and propose a heuristic algorithm. We present simulation results for the proposed approaches in a two-layer network setup and observe a significant common rate increase compared to single-layer wireless networks.
\end{abstract}

\section{Introduction}
In cellular networks, employing low-power base stations to umbrella macrocells offers more opportunities to provide increased system capacity and throughput, enhanced coverage and seamless service to reach high data rates \cite{Rappaport}. Therefore, the application of low-power base stations are proposed in recent standards such as LTE and LTE-Advanced \cite{ReleaseNine,ReleaseTen}. There are two types of low-power base stations considered in this paper and they are classified according to the existence of the backhaul connections with the high-power base station layers. First, we consider a microcell base station overlay that is connected to the macrocell layer through a fast backhaul. Second type of low-power base stations we considered here are relays that only have decode-and-forward capabilities and have no backhaul connections to the macrocell layer.

The throughput maximization problem in wireless networks has been vastly studied in the literature for various optimization objectives and constraints. In \cite{Geometric}, Chiang \emph{et al.} analyze the sum-rate maximization, the worst-case user rate maximization and specific user rate maximization objectives, and propose geometric programming (GP) and heuristic solutions depending on the signal-to-interference ratio (SIR) regime. In \cite{BinaryPC}, Gjendemsj \emph{et al.} consider the weighted sum rate maximization problem in single-layer wireless networks. Julian \emph{et al.} study the power allocation problem in single layer networks to maximize the minimum SIR with certain quality of service and fairness constraints \cite{BoydAdHoc}. Karakayali \emph{et al.} solve the common rate maximization problem in single layer networks in \cite{Karakayali}. On the other hand, there are fewer works on maximizing throughput in two-layer wireless networks. In \cite{RamanConf}, Raman \emph{et al.} propose a linear programming (LP) solution to the same problem in a two-layer network system that consists of macrocell and relay layers. Our paper differs from the aforementioned papers in two ways. We first present an analytical solution to solve the power control problem in two-layer wireless networks without any maximum power constraints and identify the necessary conditions for feasible power levels. We use this framework to define boundary points and apply LP to heuristically solve the common rate maximization problem in two-layer systems where we consider macrocell-microcell and macrocell-relay systems. For comparison, we also investigate the maximum common rate of an uncoordinated system without power control.
%In this paper, we assume a centralized power control scheme with a high computation capabilities and use it to determine and assign transmit power levels of all the base stations in the downlink. Furthermore, we investigate the common rate increase in both macrocell-microcell and macrocell-relay systems compared to single-layer networks through simulations.

\section{System Model}\label{SystemModel}
\begin{figure}[tb!]\begin{center}
  \includegraphics[height=2.3in]{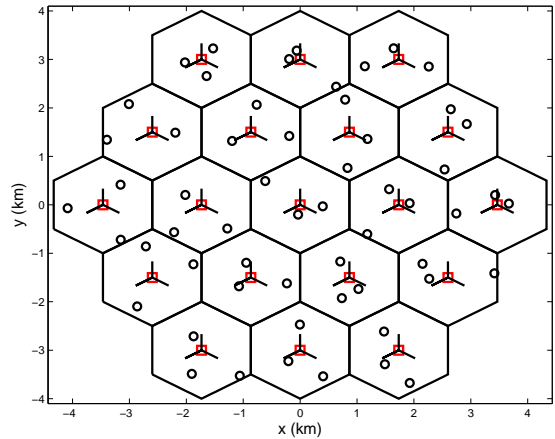}\\ %{singlelayer12}\\
  \caption{Single-layer network layout with $19$ hexagonal cells is displayed. Each cell has three-sector $120^o$  directional antennas positioned at the center of the cell and each sector has one randomly placed user. Squares and circles depict macrocell base stations and users, respectively.}\label{baseline_sector}\end{center}
\end{figure}

In this section, we present three different system models used in this paper. In all three system models, we assume a central processor with sufficiently high computational capability, as in \cite{RamanConf,Karakayali,Goodman}. First, as a reference scenario, we consider a single-layer radio network where only macrocells are present in the system. As in \cite{RamanConf,Karakayali} assume an idealized $19$-cell hexagonal layout shown in Figure \ref{baseline_sector}. In each hexagonal cell, the macrocell base stations are located at the center of the cell. Every macrocell base station is equipped with three sector antennas such that each sector covers $120^\circ$ within the cell. We consider universal frequency reuse such that all the base stations operate on the same frequency band. In order to avoid edge effects, we employ the wrap-around technique \cite{RamanConf}.

In our simulations, we place the users one-by-one randomly to each cell such that each sector only serves a single user. During the user generation process, the only condition we seek to satisfy is that each user has the lowest highest received signal strength from its associated base station and not from any of the neighboring base stations. If this condition is not satisfied and the user needs to be handed over to the neighboring base station, that user is discarded and another one is generated. This method is the same as in the ones in \cite{RamanConf,Karakayali}.

The first transmission is devoted to estimating channel parameters for all the users in which pilot-assisted channel estimation methods such as least-squares (LS) or minimum mean square error (MMSE) estimators are used. We refer the reader to \cite{channelest1,channelest2,channelest3} and references therein for details on the channel estimation methods for various systems. Since in this paper we only evaluate the performance increase with low-power base station deployment, we assume perfect channel estimation and leave the effects of the channel estimation error as future work.

We investigate a wide range of system loads and outage rates which are two important design parameters that closely affect the system throughput in wireless networks. Due to severe path losses and shadowing effects, the network performance can be significantly degraded when $100\%$ system load is considered. For this reason, network design engineers use outage percentage as a design parameter in order to guarantee a common rate to the remaining users. In this paper, we investigate a wide range of system loads from $80\%$ to $100\%$. In the procedure to decide the users in outage, we discard users one-by-one from the system based on the worst SINR values until the designed system load is reached. By this way, coupling among the users are avoided. The user discarding procedure here is similar to the ones in \cite{Karakayali,RamanConf}.

In the next transmissions, power control is applied to reduce intercell interference and the common rate maximization problem is solved by the central processor that can handle the high computational load in assigning power levels to the base stations. In the following sections, we present these techniques used here in detail. At the end of this step, the baseline system is completed.

The second and the third systems we evaluate include additional overlay of low-power base stations in order to help macrocell base stations to conserve power, increase coverage, and possibly improve system throughput. For the macrocell-microcell system, we assume that backhaul connections exist to connect both layers to the central processor where the channel gains of the users are analyzed to decide on downlink transmit powers. The macrocell-relay system includes no backhaul to connect both layers, only the macrocell layer base stations are connected among each other. The relays in the system turn on if they can decode the message from the macrocell base station they are associated with, otherwise they are turned off. Furthermore, we assume that through a control channel, the active relays notify the associated base stations about their status and the base stations send the assigned power levels to the relays in return.

\begin{figure}[tb!]\begin{center}
  \includegraphics[height=2.3in]{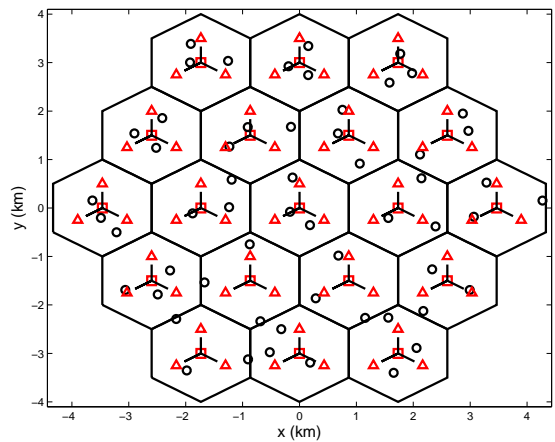}\\%{ramanlayout}\\
  \caption{Two-layer hierarchical network layout with $19$ high-power base stations overlaid with $19$ low-power base
    stations placed at predefined locations. Additional low-power base station layer employs omnidirectional antennas. Squares, triangles and circles show macrocell base stations, low-power base stations and users, respectively.}\label{twolayer_fig}\end{center}
\end{figure}

Figure \ref{twolayer_fig} depicts our two-layer cellular system layout where high-power and low-power base stations are deployed in the same area. Note that the position and the quantity of low-power base stations are very important while evaluating the system performance. In this work, we do not seek to place them in optimal locations. Instead, low-power base stations are located at a half distance from the cell center to cell edge in the boresight direction of the sector antennas.

The user generation methodology and user discarding procedures are the same as in single-layer networks. Note that the generated channel matrix is an augmented matrix compared to the baseline case such that it additionally considers parameters regarding the second layer. Once the channel conditions are learned, the central processor applies power control, solves the rate-maximization problem, assigns power levels to all the base stations and notifies them through backhaul in the case of macrocell-microcell systems or through short notification messages in the control channel for the macrocell-relay systems. In the sequel, we describe the analytical framework for the power control processes, identify necessary conditions for feasible power levels and analyze the common rate maximization problem for both single-layer and two-layer cellular networks.

%===
%the central processor discards users one-by-one and solves the power control and common rate maximization problem. Through the backhaul connections, central processor notifies both layers and updates the power levels such that the second transmission power levels are optimized to provide maximum common rate with power control. In the case of relays, only the relays that can decode the first transmission are considered as active. The central processor solves the common rate-maximization and power control problem and assigns the power levels to base stations in both layers by short notification messages to update their power levels.
%===
\section{Power Control}\label{PowerControlSection}
\subsection{Single-Layer System}
In the single-layer system, only high-power macrocell base stations are considered. First, we identify the target signal-to-interference-plus-noise-ratio (SINR) for user $i$ as \bea \gamma_i =
\frac{g_{ii}p_i}{\sum\limits_{j \neq i} g_{ij}p_j + \sigma_i^2}
\eea where $p_i$ is the transmit power of the $i$th base
station, $\sigma_i^2$ denotes the noise power, and $g_{ij}$ denotes the channel coefficient between user $i$ and base station $j$ and it includes the path loss and the shadowing. When we rearrange the terms above and divide by
$g_{ii}$, we get \bea p_i = \gamma_i \sum\limits_{j \neq i}
\frac{g_{ij}}{g_{ii}}p_j + \gamma_i \frac{\sigma_i^2}{g_{ii}}.
\eea In vector-matrix form, the above equation can be written to
include all users as follows \bea \textbf{p} = \textbf{DF}
\textbf{p} + \textbf{D} \textbf{u} \label{p_dfp_u}\eea where $ u_i = \sigma_i^2 /
g_{ii}$, $\textbf{D} = \mathrm{diag}\{\gamma_1,\ldots,\gamma_N\}$,
\begin{align} \textbf{F} = \begin{cases}g_{ij}/g_{ii} & \text{if } j \neq i \\ 0 & \text{if } j = i, \end{cases}
\end{align} and we refer to $\textbf{F}$ as the normalized channel gain matrix. The matrix $\textbf{D}$ includes the target SINR values for each user on its diagonal entries. In the case of the common-rate maximization problem, all the users target the same SINR. Therefore, $\textbf{D}$ can be reduced to a scalar $\gamma_0$ such that the Additive White Gaussian Noise (AWGN) channel capacity is $\log_2(1+\gamma_0) = r_0$. Hence, the optimum power levels for a single-layer system providing a common rate target of $r_0$ can be shown as
\bea \textbf{p}^* = (\textbf{I}_N - \gamma_0 \textbf{F})^{-1} \gamma_0
\textbf{u} \label{opt_single}\eea
where $\textbf{I}_N$ is an $N\times N$ identity matrix. In the following, we identify the nonnegativity of the assigned power levels to macrocell base stations $\textbf{p}^*$ in single layer. First, we assume that entries of channel gain matrix $\textbf{F}$ are realizations of the underlying stochastic processes, namely they are all random path loss variables such that they are all independent and $\textbf{F}$ is a full-rank matrix. Then, we apply the Perron-Frobenious theorem \cite{TheoryOfMatrices} which states that a real square matrix with nonnegative entries has a unique largest eigenvalue and its corresponding eigenvector has strictly nonnegative components. Since this theorem only applies to irreducible matrices, we need to question the irreducibility of $\textbf{F}$. Note here that a square matrix is reducible if and only it can be placed into block upper triangular form by simultaneous row-column permutations.

Following the same argument in \cite{Goodman}, $\textbf{F}$ becomes reducible if and only if there exists more than one $0$ element on one row. Since we include the channel gains of every base station to every user using the wrap-around technique, we can conclude that $\textbf{F}$ is irreducible. Hence, for an irreducible nonnegative matrix $\textbf{F}$, there always exists a positive real eigenvalue of $\textbf{F}$, $\lambda^*$ such that $\lambda^* = \max \{\lambda\}_{i=1}^N = \rho(\textbf{F})$, which is called the spectral radius of $\textbf{F}$. The eigenvector associated with this eigenvalue is element-wise nonnegative \cite{TheoryOfMatrices}. Using these results, we can rewrite (\ref{opt_single}) such that
\begin{align} \textbf{p}^* = (\textbf{I}_N - \gamma_0\textbf{F})^{-1} \gamma_0 \textbf{u} = \frac{1}{1-\gamma_0\rho(\textbf{F})} \gamma_0\textbf{u} \geq \textbf{0}\end{align}
and for the convergence of the solution, we seek that the spectral radius of $\textbf{F}$ needs to be less than $1/\gamma_0$, $\rho(\textbf{F})<1/\gamma_0$. Here, we note that this condition was already identified in \cite{ZanderCentral,Zender,Goodman,Foschini,GoodmanDPC}.
%A major drawback of this approach is that it is a centralized solution. It is impractical for a central processor to have the
%perfect knowledge of all path loss values for all users in the system and determine the appropriate power levels for every user
%and send back the power levels to the associated base stations in a reasonable time. Therefore, several distributed solutions are
%proposed where each base station can iterate and adjust its power level using only the local acquired information without the need
%for a global central processor \cite{Zender,Foschini,GoodmanDPC}. One distributed power solution is proposed by Foschini and
%Mijalcic where each base station updates its power level using the following rule \cite{Foschini} \bea p_i^{n+1} =
%\frac{\gamma_i}{g_{ii}} \left(\sum\limits_{j \neq i}g_{ij} p_j^n + \sigma_i^2 \right) \eea where $p_i^n$ denotes the power level at
%base station $i$ at $n$th iteration. In vector-matrix form, the power update rule can be written as \bea \textbf{p}^{n+1} =
%\textbf{F} \textbf{p}^{n} + \textbf{u}  \eea where $\textbf{p}^{n}$ denotes the vector of macrocell transmit power
%levels at $n$th iteration, and $\textbf{F}$ and $\textbf{u}$ are as defined above. Note that, in \cite{Foschini} it has been shown
%that for any initial power levels, the base station power levels converge exponentially to the optimal solution using the above
%power update rule.
\subsection{Two-Layer System}
In the two-layer system, we consider cross-layer interference from both macrocell and microcell
layers and update our target SINR definition such that
\bea \gamma_i = \frac{g_{ii} p_i + h_{ii} q_i}{\sum_{j \neq i}
\left(g_{ij} p_j + h_{ij} q_{j} \right)+\sigma_i^2} \eea where
$p_i$ denotes the power transmitted from the $i$th macrocell base
station and $q_i$ is the transmit power of the microcell base
station $i$. The channel coefficient representing the path loss and shadowing from the macrocell base station $j$ to user $i$ is denoted by $g_{ij}$ and for the microcell $j$ to user $i$ is denoted by $h_{ij}$. The noise power at receiver $i$ is represented as $\sigma_i^2$. Using these, the above equation can also be expressed as the following when we rearrange terms and divide every term by
$g_{ii}$
\begin{align} p_i +
\frac{h_{ii}}{g_{ii}} q_i = \gamma_i \sum_{j \neq i}
\left(\frac{g_{ij}}{g_{ii}} p_j + \frac{h_{ij}}{g_{ii}} q_{j}
\right) + \gamma_i\frac{\sigma_i^2}{g_{ii}}. \end{align} One can rewrite
the above equation in vector-matrix form as \begin{align}
\underbrace{\left[ \textbf{I}_N | \textbf{C}_{N\times N}
\right]}_{\textbf{A}_{N\times
2N}}\underbrace{\left[\bary{c}\textbf{p} \\ \textbf{q}
\eary\right]}_{\textbf{x}_{2N\times 1}} = \textbf{D} \underbrace{\left[\textbf{F}_{N\times N} | \textbf{G}_{N\times N}
\right]}_{\textbf{B}_{N\times 2N}}
\underbrace{\left[\bary{c}\textbf{p} \\ \textbf{q}
\eary\right]}_{\textbf{x}_{2N\times 1}}  + \textbf{D}
\underbrace{\left[\bary{c} \frac{\sigma_1^2}{g_{11}} \\ \vdots \\ \frac{\sigma_N^2}{g_{NN}} \eary\right]}_{\textbf{u}_{N
\times 1}} \nonumber\end{align} where $N$ denotes the number of users in
the system, $\textbf{I}_N$ is an $N\times N$ identity matrix and
$\textbf{C}$ and $\textbf{G}$ matrices are as shown below
\begin{align} \textbf{C} &= \begin{cases} \frac{h_{ii}}{g_{ii}} & \text{if } j = i
\\ 0 & \text{if } j \neq i \end{cases}, \hspace{5mm} \textbf{G}  = \begin{cases}
\frac{h_{ij}}{g_{ii}} & \text{if } j \neq i \\ 0 &
\text{if } j = i.
\end{cases}\end{align} We refer to $\textbf{G}$ as the normalized channel gain matrix for the low-power base station
layer where the normalization is carried out with respect to macrocell layer path loss values, $g_{ii}$.

Similar to the single-layer case, for the common-rate maximization problem, $\textbf{D}$ can be reduced into a scalar $\gamma_0$ and to determine the optimum power levels for the two-layer system, we need to solve $\textbf{A} \textbf{x} = \gamma_0 \textbf{B} \textbf{x} +
\gamma_0 \textbf{u}$. When we rearrange the terms on each side, the power control problem can
be expressed as
\begin{align} \textbf{A} \left(\textbf{I}_{2N} - \gamma_0 \widetilde{\textbf{B}} \right) \textbf{x} = \gamma_0 \textbf{u} \end{align}
where $\widetilde{\textbf{B}} =
\textbf{A}^{-1} \textbf{B}$ such that $\textbf{B}  = \textbf{A}\widetilde{\textbf{B}}$ and
$\textbf{A}^{-1}$ denotes the adjoint matrix of the rectangular matrix $\textbf{A}$. Then, the
optimal solution for the two-layer cellular system becomes
\begin{align} \label{opt_two} \textbf{x}^* = \gamma_0 (\textbf{I}_{2N} - \gamma_0 \widetilde{\textbf{B}})^{-1} \textbf{A}^{-1} \textbf{u}.
\end{align}
To analyze the existence and nonnegativity of the optimal solution vector $\textbf{x}^*$, we follow a similar analysis as
in the single-layer case and apply the Perron-Frobenious theorem. Then, we see that there always exists a componentwise
nonnegative power vector $\textbf{x}^*$ as long as the spectral radius of $\widetilde{\textbf{B}}$ is less than $1/\gamma_0$. Hence, we conclude that the necessary condition for the above solution always to yield feasible power levels for two-layer cellular systems is $\rho(\widetilde{\textbf{B}})<1/\gamma_0$.

\section{LP Solution and Heuristic Common Rate Maximization Algorithms}\label{LPTitle}
In this section, we consider maximum power constraints for each base station due to the physical limitations of radio amplifiers in the base stations and introduce our common rate maximization algorithm. In cases where the analytical solutions obtained using (\ref{opt_single}) or (\ref{opt_two}) exceed these levels, we need to pursue a different approach. Raman \emph{et al.} have proposed a linear programming solution to this problem in \cite{RamanConf} where the maximum power level constraints are introduced to the common rate maximization problem. Note that the two-layer system considered in \cite{RamanConf} is our third system model where macrocell cellular network is overlaid with a relay layer.

In what follows, we introduce the common rate maximization problem in two-layer cellular networks and propose our heuristic solution. The common rate maximization problem for the macrocell-microcell system can be stated as
\begin{align}\label{lpsol}\begin{array}{rl}  \max\limits_{\small{\textbf{p,q}}}  & r_0 \\
  \mathrm{s.t.} & \log_2 \small{\left(1 + \frac{g_{ii} p_i + h_{ii} q_i}{\sum_{j \neq i} \left( g_{ij} p_j + h_{ij} q_{j}   \right)+\sigma_i^2} \right)} \geq r_0, \hspace{.05in} \forall i \\
    & 0 \leq p_i \leq p_{\max}, \hspace{.05in} \forall i \\ & 0 \leq q_i \leq q_{\max}, \hspace{.05in} \forall i
\end{array}\end{align}
where $r_0 = \log_2(1+\gamma_0)$ bits/sec/Hz denotes the common rate provided to the users in the system, and $p_{\max}$ and $q_{\max}$ are the maximum transmit power levels of macrocells and microcells, respectively. The first constraint ensures that the users get at least their target SINR as in \cite{RamanConf,Geometric} and the last two constraints impose the maximum power level constraints. The maximization problem solution in (\ref{lpsol}) yields a new set of power levels considering the maximum power constraints in each layer.

For the macrocell-relay system only those relays that can fully decode the first transmission are included in the solution. Then, we update the power constraints in (\ref{lpsol}) as %for this system as
\begin{align}\label{lpsolrelay}
    \begin{array}{rl}
     & 0 \leq p_i \leq p_{\max}, \hspace{.05in} \forall i \\ & 0 \leq q_i \leq q_{\max}, \hspace{.05in} i \in \mathcal{S} \\ & q_i = 0,\hspace{.19in} \hspace{.05in} i \in \mathcal{S}^c
 \end{array}
\end{align}
where $\mathcal{S}$ denotes the set of relays that can decode the first macrocell base station transmission and $\mathcal{S}^c$ denotes its complement set.

Although we state the common-rate maximization problem in (\ref{lpsol}) and (\ref{lpsolrelay}), the proposed heuristic common rate maximization algorithm solves the following objective to employ power control
\begin{align}\label{lpsolsum}
\hspace{-.1in}     \begin{array}{rl}
     \min\limits_{\textbf{p,q}}  &  \textbf{1}^T\left[\begin{array}{c}\textbf{p}\\ \textbf{q}\end{array}\right] \end{array}
\end{align}
with the associated constraints in (\ref{lpsol}) or (\ref{lpsolrelay}). In the first step, we target a small but a feasible common rate $r_0$ and increase it in small $\Delta r$ increments until the maximum power constraints are violated in either layer. The reason that we use the heuristic algorithm rather than the analytical solution is that the analytical solution reduces the macrocell power levels and increases the transmit power levels for the low-power base station above the permissible power levels since the users are typically closer to low-power base stations on average due to the inherent geometry. For this reason, we employed the LP solution in (\ref{lpsolsum}) to solve the common rate maximization problem under the maximum power constraints. %Note that, in the heuristic algorithm, power control is also applied.
\begin{algorithm}
  \caption{Heuristic Algorithm for Single-Layer Networks}\label{singlelayerlpalgorithm}
  \begin{algorithmic}[1]
      \While{$\gamma_0 < 1/\rho(F)$}
        \State $\textbf{p} = (\textbf{I}_{N} - \gamma_0 \textbf{F})^{-1} \gamma_0 \textbf{u}$;
        \If{$p_i > p_{\max}$, $i \in \{1,\ldots,N\}$}
             \State $r_0 = r_0 - \Delta r$; $\textbf{p} = (\textbf{I}_{N} - \gamma_0 \textbf{F})^{-1} \textbf{u}$; \textbf{Exit}
        \Else
            \State $r_0 = r_0 + \Delta r$;
        \EndIf
      \EndWhile\label{euclidendwhile}
      \State \textbf{return} $r_0$,$\textbf{p}^*$
  \end{algorithmic}
\end{algorithm}
\begin{algorithm}
  \caption{Heuristic Algorithm for Two-Layer Networks}\label{twolayerlpalgorithm}
  \begin{algorithmic}[1]
      \While{$\gamma_1 < 1/\rho(\widetilde{B})$}
        \State Solve (\ref{lpsolsum}) with the constraints in (\ref{lpsol}) or (\ref{lpsolrelay})
        \If{$p_i > p_{\max}$ $\vee$ $q_i > q_{\max}$,  $i \in \{1,\ldots,N\}$}
             \State $r_1 = r_1 - \Delta r$
             \State Solve (\ref{lpsolsum}) with the constraints in (\ref{lpsol}) or (\ref{lpsolrelay})
             \State \textbf{Exit}
        \Else
            \State $r_1 = r_1 + \Delta r$;
        \EndIf
      \EndWhile\label{euclidendwhile}
      \State \textbf{return} $r_1$,$\textbf{p}^*$,$\textbf{q}^*$
  \end{algorithmic}
\end{algorithm}

Algorithms \ref{singlelayerlpalgorithm} and \ref{twolayerlpalgorithm} are used to solve the heuristic common rate maximization problem in single-layer networks and two-layer networks. In the latter algorithm, we solve the objective in (\ref{lpsolsum}) with the convex constraints in (\ref{lpsol}) for the macrocell-microcell system and use (\ref{lpsolsum}) with the nonconvex power constraints in (\ref{lpsolrelay}) for the macrocell-relay system solution. For comparison purposes, we record the maximum common rates $r_0$ and $r_1$ at the end of each while loop. The solution converges for both cases since we generate an increasing sequence of rates that are bounded \cite{RamanConf}.

\section{Simulations}\label{Simulations}
In this section, we present our simulation results for the common rate maximization problem when two-layer cellular systems are considered.
We follow the simulation setup described in Section \ref{SystemModel}, and without loss of generality, consider the $19$ hexagonal cell layout in Figure \ref{baseline_sector}. In our simulations, we consider macrocell base stations with three sector antennas, each covering $120^o$ within the cell. We assume an idealized cell radius of $1$ km and employ the wrap-around technique to avoid the edge effects. The horizontal radiation pattern used for the three-sector antenna is
\begin{align} A(\theta) = - \min \left(12 \left( \frac{\theta}{\theta_{3\mathrm{dB}}}\right)^2, \hspace{.05in} A_{\max}\right) \hspace{.05in} (\mathrm{dBi}) \end{align}
where $-180 \leq \theta \leq 180$, $\theta_{3\mathrm{dB}}$ and
$A_{\max}$ denote the $3$ dB beam width and the maximum
attenuation, respectively, and they are taken as
$\theta_{3\mathrm{dB}}=65^o$ and $A_{\max} = 20$ dB \cite{UmtsCostHataWI}. For the microcell and relay base stations, we consider only omnidirectional antennas.

Following the user placement procedure described in Section \ref{SystemModel}, $57$ users are placed in the system such that they share the same resource. In the first transmission, the channel estimation for all users is carried out and the central processor forms the channel gain matrix. Based on this information, the system loading is adjusted. In our simulations, we sweep a wide range of system loading $80\%$ to $100\%$.

We consider different channel models to model propagation in macrocell and microcell environments while adopting the proposed parameters specified in \cite{UmtsCostHataWI}. The path loss model for urban macrocells is based on modified COST $231$ Hata urban propagation model and the microcell non-line of sight (NLOS) environment is based on COST $231$ Walfish-Ikegami NLOS model. Furthermore, in our simulations, we assume the macrocell base stations height as $h_{BS} = 32$ m, mobile height as $h_{MS} = 1.5$ m, and carrier frequency as $f_c = 1900$ MHz. Then, the macrocell path loss model becomes
\begin{align} PL_{\mathrm{Macro}}(\mathrm{dB}) = 34.5 + 35 \log_{\hspace{.01in}10}\hspace{.05in}(d/m)\end{align}
and we consider log-normal shadowing with a standard deviation of $8$ dB. For NLOS microcell path loss model, we take microcell base station height as $h_{BS} = 12.5$ m, average building height as $12$ m, mobile height as $1.5$ m, orientation for all paths as $\phi = 30^\circ$, building to building separation as $50$ m and street widths as $25$ m. The resulting path loss model for NLOS microcells at $f_c = 1900$ MHz is
\begin{align} PL_{\mathrm{Micro}}(\mathrm{dB}) = 34.53 + 38 \log_{\hspace{.01in}10}\hspace{.05in}(d/m) \end{align}
and log-normal shadowing with a standard deviation of $10$ dB.

%in order to make a fair comparison to the results in \cite{RamanConf}, we take the same maximum base station power levels and antenna gains. We assume the maximum transmission power levels of macrocell base stations as $30$ W and low-power base stations as $1$ W.
Moreover, we assume the maximum downlink power levels for macrocell base stations, microcell base stations and relays are $43$ dBm, $33$ dBm and $30$ dBm as in \cite{ReleaseNine,RamanConf}. Also, the base station transmitter antenna gain, the user receiver antenna gain and the other losses such as cabling losses are taken as $15$ dB, $-1$ dB and $-10$ dB, respectively. A major difference between our paper and \cite{RamanConf} is that the same path loss model is used to model both macrocell and relay environments in \cite{RamanConf}  and it is given by
\begin{align} PL(\mathrm{dB}) = 31.5 + 38 \log_{\hspace{.01in}10}\hspace{.05in}(d/m). \label{ramanpathlossmodel}\end{align}
where the log-normal shadowing parameter has a standard deviation of $8$ dB. We note here that the transmission power levels of the base stations bound the permissible power level ranges and the antenna gains determine the received power levels along with the path loss models used. Hence, these parameters closely affect the maximum achievable common rate.

\begin{figure}[tb!]\begin{center}
  \includegraphics[height=2.5in,width=3.25in]{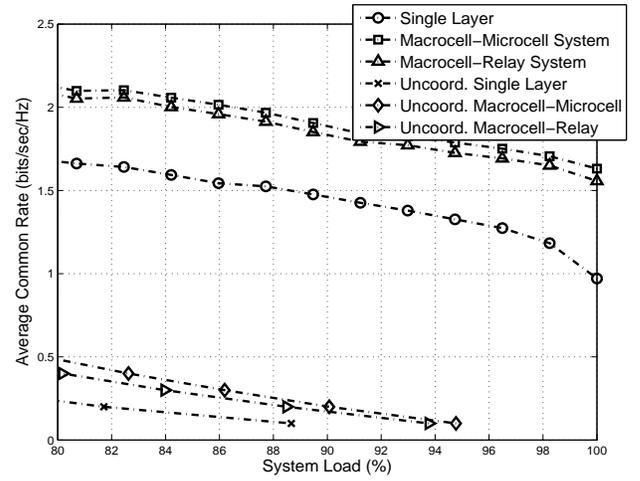}%{article2_maxrate_ltemodel_500iters_3to10}\\
  \caption{The figure shows the average common rate versus system loading in a $57$ user system when COST $231$ Hata urban propagation model and the COST $231$ Walfish-Ikegami NLOS model are used.}\label{lterate}\end{center}
\end{figure}

In our simulations, we used $r_0 = 0.1$ bits/sec/Hz and step size $\Delta r = 0.1$ bits/sec/Hz. Then, the heuristic algorithms for the common-rate maximization are applied. In Fig. \ref{lterate}, we plot the maximum common rate versus various system loads when the COST $231$ Hata urban propagation model and the COST $231$ Walfish-Ikegami NLOS model are used. The first observation we make is that by employing low-power base stations to support high-power macrocell base stations brings $23.52\%$ to $68.03\%$ increase in common rate throughput compared to the single layer system. For instance, when $90\%$ system load is considered, the maximum common rate increased from $1.48$ bps/Hz to $1.85$ bps/Hz ($25.45\%$ increase) and to $1.91$ bps/Hz ($29.14\%$ increase) for macrocell-microcell and macrocell-relay systems, respectively.

Second, as the system load increases, the maximum common rate decreases due to the increase in both intercell and intracell interference. Obviously, when the system load is reduced by the central processor, the outage in the system increases and the availability of the service decreases. Hence, the maximum allowable system load in the system is a design parameter for the network design engineers to trade-off between the maximum common rate and the outage of the system. This parameter can also be used in admission control to make sure a certain level of common rate is offered to the users at all times.

\begin{figure}[tb!]\begin{center}
  \includegraphics[height=2.5in,width=3.25in]{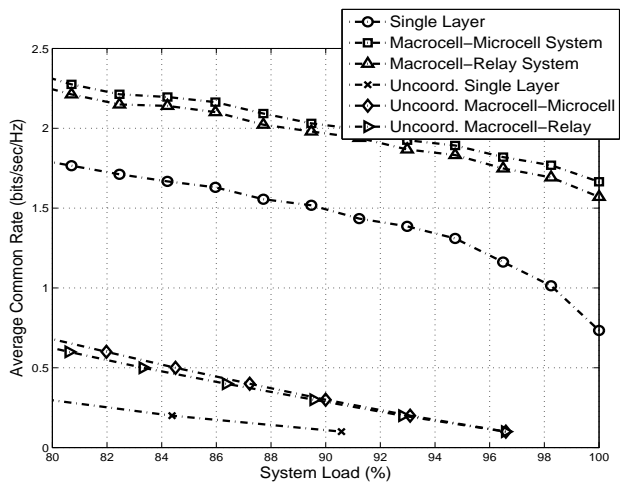}%{article2_maxrate_ramanmodel_500iters_3to10}\\
  \caption{The figure shows the average common rate versus system loading in a $57$ user system when (\ref{ramanpathlossmodel}) is used to model propagation in both layers.}\label{ramanrate}\end{center}
\end{figure}

Third, we see that the macrocell-microcell system offers slightly more common rate compared to the macrocell-relay system. In the former system, microcells are connected to the macrocell layer through backhaul, whereas in the latter, the relays need to be able to decode the message from the macrocell base stations. In our simulations, we observe that on the average $35$ relays are active. In cases where the relays cannot decode the macrocell transmission, they are not included in the solution and this clearly reduces the degrees of freedom in the solution.

Also, for comparison purposes, we simulated the uncoordinated transmission where all the base stations transmit at full power instead of implementing power control. This is to observe the tradeoff between complexity and performance increase that power control brings. We see that even in the uncoordinated network system, employing low-power base station overlay to macrocell system improves the common rate performance. As an example, at $85\%$ system load, the uncoordinated single layer system consisting of only macrocell base stations can only provide up to $0.15$ bps/Hz and this rate increases to $0.33$ bps/Hz for the macrocell-microcell and to $0.28$ bps/Hz for the macrocell-relay systems. We clearly see that almost half an order of magnitude increase in common rate can be achieved when power control is employed. An important observation is that the system load cannot exceed $95\%$ without coordination in all three systems.

Figure \ref{ramanrate} depicts the results for the same analysis when the path loss model in (\ref{ramanpathlossmodel}) is used to model the propagation loss in both layers. We observed that the macrocell-microcell and macrocell-relay networks provided $28.87\%$ to $127.09\%$ and $25.32\%$ to $114.16\%$ common rate increase, respectively. For instance, at $90\%$ system load, the maximum common rate increased from $1.52$ bps/Hz to $2.03$ bps/Hz ($33.83\%$ increase) and to $1.98$ bps/Hz ($30.5\%$ increase) for the macrocell-microcell and macrocell-relay systems, respectively. We see that our simulation results and the results presented in \cite{RamanConf} are consistent.
\section{Conclusion}\label{Conclusions}
The deployment of low-power base station overlay to high-power base station layers offer increased common rate throughput in the cellular radio systems. These additional low-power base stations offer more opportunities to hand over the transmissions from macrocells to low-power layers where the same data rates can be achieved with less transmit power in the downlink. This advantage reduces the interference, brings significant power savings to the operators and provides solutions to the coverage problems. In this paper, we presented an analytical solution framework to solve the power control problem in two-layer cellular networks and outlined the feasibility conditions for determining the power levels. We proposed a heuristic solution to maximize the common rate offered in the two-layer systems. Through simulations, we showed that significant increase in common rate can be achieved for macrocell-microcell and macrocell-relay systems.

\end{document}